\documentclass[aps,pre,reprint]{revtex4-2}

\usepackage{graphicx}

\begin{document}
\title{Shock width measured under liquid and solid conditions in a 2D dusty plasma}
\author{Anton\ Kananovich}
\author{J.\ Goree}
\affiliation{Department of Physics and Astronomy, University of Iowa, Iowa City, Iowa 52242, USA}
\date{\today}
\begin{abstract}
Widths of shocks are compared, under liquid and solid conditions, for a two-dimensional layer of charged microspheres levitated in a plasma. In this strongly coupled dusty plasma, a shock was launched as a blast wave by moving an exciter wire at a supersonic speed and then bringing it to a halt. Runs were repeated with the layer of microspheres prepared two ways:  a crystalline-like solid, and a liquid. The liquid was sustained using laser heating, with conditions that were otherwise the same as in the solid.
The shock width was found to be less in a liquid than in a solid, where it was 4 to 6 lattice constants. These measurements were based on the high-gradient region of density profiles. The profiles were obtained from particle coordinates, measured by high-speed video imaging. The spatial resolution was improved by combining particle coordinates, in the shock's frame of reference, from a sequence of images.
\end{abstract}
\maketitle

\section{\label{secIntro}Introduction}

For shocks in all kinds of substances, the structure of a shock and in particular its width have attracted scientific interest for many years~\cite{talbot1962survey,sakurai1957note,liepmann1966theoretical,hoover1979structure,holian1980shock,hansen1960thickness,anisimov1997shock,anand2016effects}. Although a shock is often described as a discontinuity in parameters such as number density, a true discontinuity is impossible. A shock must have a finite width in the presence of collisions, as described in a gas for example by a finite mean-free-path and dissipation by viscous effects~\cite{zeldovich2002physics}. Most studies on this topic relied on either analytical theory or simulation. There seems to be a paucity of experiments, which can be explained by the challenge of measuring a shock profile in conventional solids, liquids and gases, where shocks propagate at speeds of the order of $10^2$ to $10^3$~m/s, and a shock width can be as small as $10^{-10}$~m.
 
These difficulties of high speed and microscopic thinness are avoided by experimenting with dusty plasmas, where typical shock speeds are of the order of centimeters per second, and shock widths have been observed to be of the order of millimeters~\cite{heinrich2009laboratory,jaiswal2016experimental,usachev2014externally}. The video microscopy diagnostics that are commonly used for laboratory dusty plasmas allow the experimenter to observe the sample at the microscopic level, tracking individual particles, and making time-resolved in-situ measurement profiles of useful quantities, such as number density. These advantages have led to many studies in the literature for dusty plasmas~\cite{marciante2017thermodynamic,pustylnik2004excitation,heinrich2009laboratory,jaiswal2016experimental,sharma2016observation,oxtoby2012tracking,oxtoby2011visualizing,samsonov2008high,samsonov2005shock,samsonov1999mach,oxtoby2013ideal,samsonov2004shock}.

The microparticles in a dusty plasma attain a large negative charge by collecting more electrons than ions. Due to this large charge, it is possible to levitate the microparticles in a layer, so that they touch no solid surface. Moreover, their large charges can cause the microparticles to interact among themselves as a strongly coupled plasmas, similar to ions in warm dense matter~\cite{doe2016report}. For this reason, a study of shock structure in dusty plasma can also potentially provide insights into warm dense matter, as well as other strongly coupled plasmas. Many kinds of strongly coupled plasmas can sustain shocks, but it is in dusty plasmas that the shocks can best be observed microscopically, using video imaging.

Besides video imaging, another laboratory method available for dusty plasma experimenters is laser heating. A radiation pressure force is applied by a laser beam, which is rastered in both the $x$ and $y$ axes, so that as a beam sweeps by, it kicks individual particles, which then collide with their neighbors. In this way, the kinetic temperature can be raised to a controlled level. The experimenter can choose to maintain either a solid phase by applying no heating, or a liquid phase at a temperature that can be varied by the choice of laser intensity, while keeping virtually all other parameters the same.

In this paper we seek to answer two questions about shocks in strongly coupled dusty plasmas.

First, \emph{we ask how the shock widths compare, in a liquid vs a solid}, for a strongly coupled dusty plasma. One could hypothesize that the widths should be different in a solid and liquid, because the finite width of shock fronts is often attributed to dissipation~\cite{zeldovich2002physics}, and the energy dissipation mechanisms can be different for liquids and solids due to viscous dissipation in a liquid, and plastic deformation or melting in a solid. Solids tend to restore their form under stress, while liquids cannot. 

That hypothesis, that the width should not be the same for shocks in liquids and solids, is not supported by our examination of the literature for a theoretical substance, a condensed matter with a Lenard-Jones potential. Density profiles for a shocked liquid~\cite{anisimov1997shock} and solid~\cite{zhakhovskii1999shock} were reported in separate papers, by related authors. Our comparison of their results suggests that the shock width is similar, for their Lenard-Jones simulations. Other than that comparison, however, our literature search revealed little to quantify the shock width in a liquid vs a solid, suggesting a need for a close comparison, especially one based on an experiment. There are several challenges for performing such an experiment, which we are able to meet using a dusty plasma, with a method of improving the spatial resolution. A fine spatial resolution  is needed because  the shock width is already a small quantity, so that detecting a small difference in shock widths is aided by a higher resolution than in previous measurements of shock widths in dusty plasmas~\cite{heinrich2009laboratory,usachev2014externally,jaiswal2016experimental}.

Second, \emph{we ask whether the hydrodynamic description is applicable for describing the shock width}. This description is commonly used in the gas-dynamic literature. It was developed for a neutral gas~\cite{liepmann2001elements,zeldovich2002physics,granger1985fluidshockwidth}, and simulations have extended is its applicability to dense liquids~\cite{hoover1979structure}. We ask whether it is also applicable to shocks in a strongly coupled dusty plasmas, or whether the large interparticle spacing in this medium will cause this hydrodynamic description to fail.  

\section{Apparatus and experimental conditions}
\label{sec_apparatus}
Our experiment draws on the methods we developed in two recent papers. In Ref.~\cite{kananovich2020experimental}, we presented a method of using a motor to propel (at a supersonic speed) a horizontal wire into a vacuum chamber, repelling microparticles and thereby launching a shock wave. In that first paper, the exciter wire was moved continuously. We analyzed the results to obtain the shock speed's dependence on the exciter speed. In Ref.~\cite{kananovich2020shocks}, we used a similar motorized exciter wire, but in a different chamber, and more importantly with a different motion. We abruptly halted the motion of the wire, so that the mechanical energy input suddenly ceased. In this second paper, we analyzed the amplitude of the shock's pulse, and found that it decayed much more slowly than can be explained by gas friction alone, indicating another energy source. Both of those papers involved shocks in a 2D layer of microspheres, which had a crystalline-like solid structure before the shock arrived.

In the present paper, we analyze the spatial profile of the shock structure. The data we analyze come from runs of the same experiment as in Ref.~\cite{kananovich2020shocks}. The experiment included six runs with shocks in solid-like conditions, and we will analyze the three of those with the weakest shock conditions. We will also analyze three runs with liquid-like conditions, which we have not previously reported. 

The data from this recent experiment~\cite{kananovich2020shocks} are suitable for the two questions we seek to answer, about the shock width. The collection of microparticles was prepared under the conditions of liquid and solid, with otherwise identical parameters. Their steady-state properties (before being shocked) were well characterized. The imaging with the top-view camera was performed with a frame rate and resolution that allow precise measurements of the shock width. The shock width was not measured in Ref.~\cite{kananovich2020shocks}, and neither was a comparison of liquid and solid conditions presented. In Ref.~\cite{kananovich2020shocks} we analyzed only data from experimental runs where the microparticle cloud was in the crystalline (solid) condition.

\begin{figure}[ht]
	\centering
	\includegraphics[width=1.0\columnwidth]{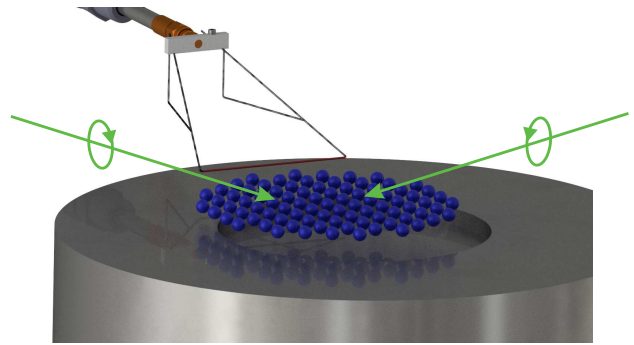}
	\caption{Apparatus. A microparticle cloud (shown not to scale) is levitated above a depression in the lower electrode. A shock is generated by the supersonic motion of a thin horizontal wire at a height 1.4~mm above the microparticles. The wire was propelled toward the center of the cloud by a motor-driven shaft, as seen at the upper left of this sketch. Independently from this shock manipulation, the kinetic temperature of the microparticle cloud could be increased by rastered laser heating with two beams, shown schematically in green. The plasma chamber, top-view camera and side-view camera are not shown.}
	\label{fWidthSetup} 
\end{figure}

\begin{figure*}[ht]
	\centering
	\includegraphics[width=1.0\textwidth]{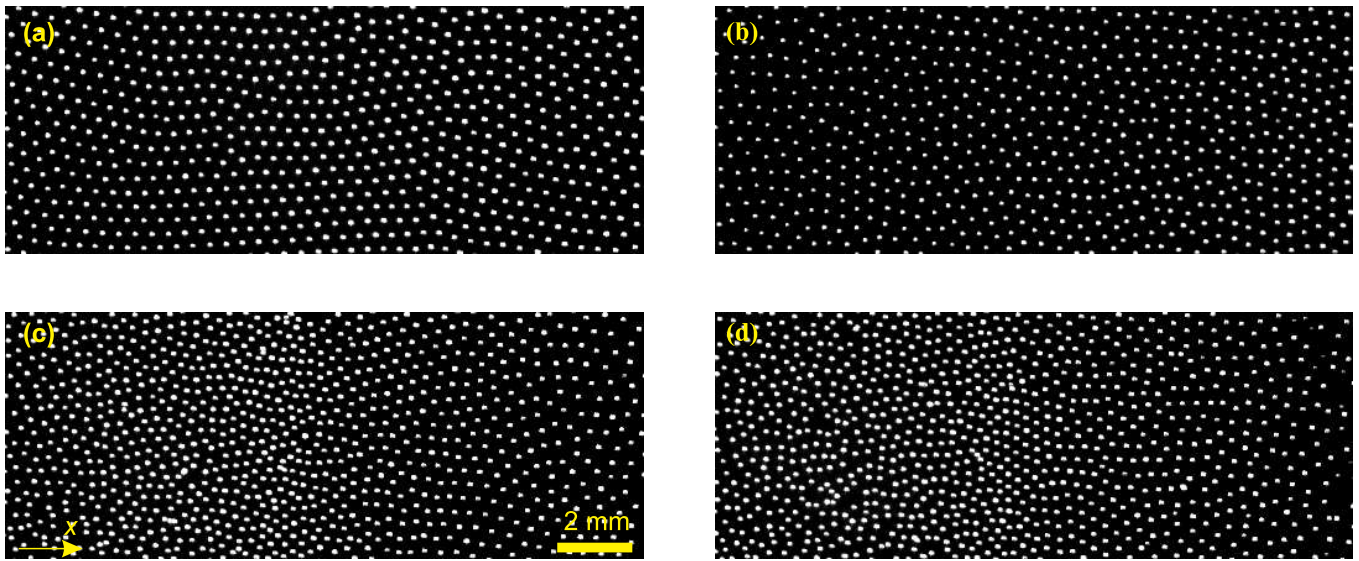}
	\caption{Top-view camera images of the microparticle cloud in these conditions: \emph{unshocked}  (a) crystalline and (b) liquid; and \emph{shocked} (c) crystalline and (d) liquid. For the unshocked crystalline condition (a), the crystal was triangular with six-fold symmetry, and the lattice constant was $b = 0.46$~mm. For the liquid conditions, rastered laser heating was applied steadily (before and during the shock manipulation) to sustain a steady elevated temperature. Each image shown here is from a single frame of a high-speed video.}
	\label{fWidthShockedUnshocked} 
\end{figure*}

The experimental setup is sketched in Fig.~\ref{fWidthSetup}. A capacitively coupled radio-frequency discharge plasma was sustained in a vacuum chamber with argon gas at a pressure of 17.0~mTorr. Polymer microparticles of 8.69~$\mu$m diameter were dropped from the top of the chamber, and the microparticles became charged so that and they were electrically levitated in the sheath above the horizontal lower electrode. The cloud of microparticles self organized naturally into a crystalline solid structure with a hexagonal structure, as shown in Fig.~\ref{fWidthShockedUnshocked}~(a). Images were recorded by a top-view video camera. For the runs with the shocks, the camera was operated at a frame rate of $R_f = 800 $~frames/second.

For our runs under liquid conditions, we applied heating~\cite{wolter2005laser,nosenko2006laser,nosenko20092d,schella2011melting,feng2012energy,schablinski2012laser,haralson2016laser,haralson2018dusty,wong2018multiple} using two laser beams that were oppositely directed onto the microsphere layer at a grazing angle. They were rastered in arcs using the method of \citeauthor{haralson2016laser}~\cite{haralson2016laser} to raise the kinetic temperature of the microspheres. With this laser heating, the collection of microparticles had a more disordered structure, as seen in Fig.~\ref{fWidthShockedUnshocked}~(b). The kinetic temperature of the microparticles, obtained from the mean square velocity fluctuation, including motion in both the $x$ and $y$ directions, was $1.7 \times 10^5$~K. (This kinetic temperature for the microparticle motion is not the same as the internal temperature of the polymer material within the microparticle, which was much cooler.) 

Before applying this laser heating, we allowed a crystal to form as in Fig.~\ref{fWidthShockedUnshocked}~(a). The areal number density of this undisturbed crystal was measured, by counting particles, as $5.7$~mm${}^{-2}$, and from that value we obtained a lattice constant $b = 0.46$~mm and a 2D Wigner-Seitz radius of $0.24$~mm. We analyzed the phonon spectrum of the undisturbed crystal, using the method of Ref.~\cite{kananovich2020shocks}, to obtain the microparticle charge $- 1.5 \times 10^4$~$e$, and screening length $\kappa =1.9$. The dust plasma frequency was dust $\omega_{pd} = 122$~s${}^{-1}$. The longitudinal sound speed $c_l$ was 16~mm/s, as explained in Ref.~\cite{kananovich2020shocks}. Under these crystalline conditions the kinetic temperature was $1.2 \times 10^3$~K and the Coulomb coupling parameter was $\Gamma = 1.40 \times 10^4$, as compared to $\Gamma = 96$ in the liquid.

The kinematic viscosity $\nu$ can be obtained for our conditions using the experimental results of~\cite{haralson2016temperature}. The experiment in~\cite{haralson2016temperature} was performed using the same chamber and the same laser-heating method as ours. In particular, we rely on Eq. (14) of~\cite{haralson2016temperature}, which is a straight-line fit to experimentally obtained data for $\nu$ vs $\Gamma$.  In that equation, the kinematic viscosity is normalized by the Wigner-Seitz radius and the dust plasma frequency. Since we measured those two quantities, along with $\Gamma$, we can estimate the kinematic viscosity for our experiment. The value obtained this way is $\nu = 1.86 $~mm${}^2$/s, which is consistent also with results from earlier experiments~\cite{nosenko2004shear,feng2012observation}.

The apparatus for exciting the shock was an electrically floating wire, oriented horizontally, and propelled in the $+x$-direction. For this experiment, the exciter wire's motion was abruptly stopped. The main data we will report in this paper were recorded after the wire had stopped, so that the shock wave was propagating without any other external manipulation that might change the conditions of the plasma. Example images of microparticles, when they are compressed by the shock, are seen in Fig.~\ref{fWidthShockedUnshocked}~(c) and (d) for a solid run and a liquid run, respectively. The mechanical configuration for the wire is sketched in Fig.~\ref{fWidthSetup}. The experimental apparatus and conditions are described in greater detail in Ref.~\cite{kananovich2020shocks}.

\section{Analysis method}

\begin{figure}[ht]
	\centering
	\includegraphics[width=1.0\columnwidth]{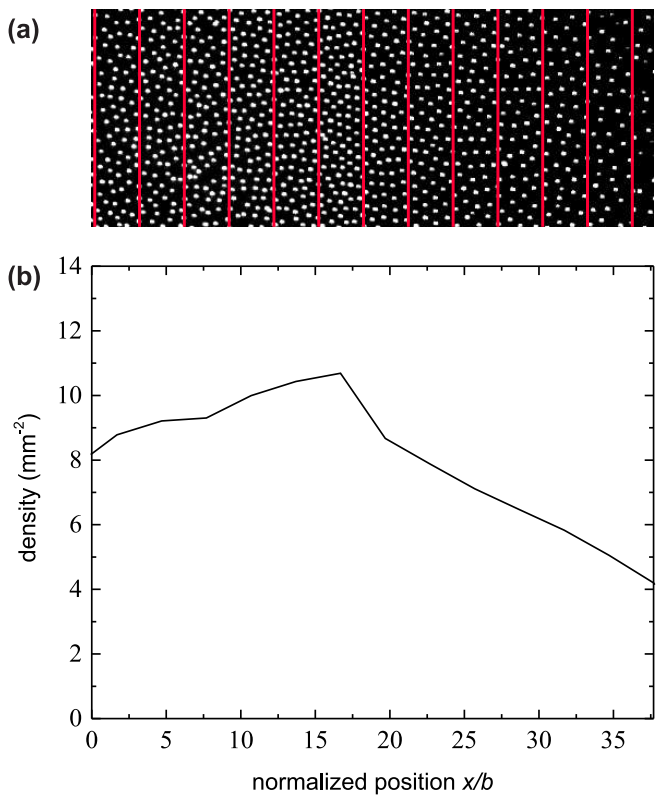}
	\caption{Snapshot of a shock propagating in the $+x$ direction. (a) Region of interest (ROI) of dimensions $17.32 \times 6.71$~mm from a top-view camera image, with lines drawn to indicate the boundaries of wide bins of width $\Delta x = 3\,b$. (b) Profile of areal number density based on the single image shown in (a). Bins of this width provide a spatial resolution too poor to allow measuring the shock width, but still useful for measuring the shock speed.}
	\label{fWidthBinning} 
\end{figure}

For each run, we analyzed image data for a sequence of video frames. A video frame is a bit-map image, and within it we selected a region of interest (ROI) $17.32 \times 6.71$~mm, as shown in Fig.~\ref{fWidthBinning}. Within this ROI, we obtained the $x$-$y$ coordinates of each microparticle, using the moment method~\cite{feng2007accurate,schneider2012nih}. In this way, we recorded positions of microparticles for each frame.

The particle-level description was converted into a density-profile description by using a binning procedure. This involved dividing the ROI into narrow rectangular regions, or bins, and counting the microparticles in each. To smooth the data, we used the cloud-in-cell method~\cite{birdsall1985plasma}, which divides the particle into the two nearest bins, weighted according to the distance of the particle from the boundary that divides the two bins. For example, if a particle is near the center of a bin, almost all of its weight will be assigned to that bin, but if it is near the boundary between bins the weight will be divided almost equally between them. This cloud-in-cell approach avoids noise generated, for example, as a microparticle moves across a bin boundary from one frame to another.

\begin{figure}[ht]
	\centering
	\includegraphics[width=1.0\columnwidth]{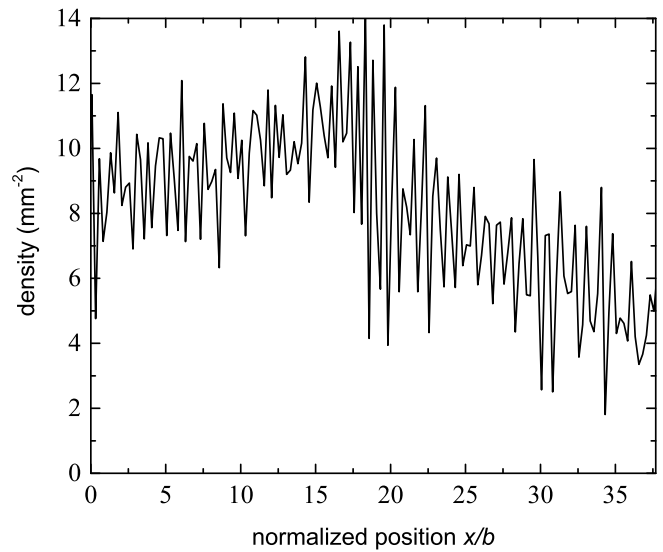}
	\caption{Profile of areal number density, for narrow bins of width $\Delta x = 0.25\,b$, for the same single image as in Fig.~\ref{fWidthBinning}. While the spatial resolution is improved by using this narrow bin width, the noise is too great to measure the shock width.}
	\label{fWidthProfileExample025b} 
\end{figure}

There is a tradeoff that must be made in choosing a bin width $\Delta x$. If the bin width is wide, for example $\Delta x = 3\,b$ as in Fig.~\ref{fWidthBinning}, the density profile will have a reduced spatial resolution. On the other hand, if the bin width is narrow, for example $0.25\,b$ as in Fig.~\ref{fWidthProfileExample025b}, the profile will be noisy.

To improve the spatial resolution while also reducing noise, we combined particle data from 25 video frames in the shock frame of reference, and we used a narrow bin width. For this purpose, the bin width was $\Delta x = 0.115$~mm~=~0.25~$b$. We were able to combine 25 frames by exploiting the steady speed of the shock. To do this, we carried out a Galilean transformation of the coordinates of each microparticle within the ROI, so that its $x$-coordinate was shifted between consecutive frames by a distance $v_{\rm{shock}} / R_f$ , where $R_f$  is the camera's frame rate. In this way, we transformed the particle position data from the laboratory frame to the frame of the shock. After this transformation, we were able to treat microparticles from the 25 consecutive frames as if they were in the same frame, thus increasing by 25-fold the number of microparticles per bin in the binning process. We chose a 25-frame time interval because during that brief time the shock moves only 1.5~mm. To judge how small this distance is, we note that it is a small multiple of the lattice constant $b = 0.46$~mm. It is also much less than the overall distance travelled by the shock, approximately 35~mm (the overall dimension of the microparticle cloud) so that the shock's spatial profile remained steady during this time interval.

The Galilean transformation in this process requires an accurate measurement of the shock's speed, $v_{\rm{shock}}$, in the laboratory frame. We obtained this value by fitting a straight line to a plot of shock position versus time \footnote{The shock speed measurement relies on a time series of measurements of the shock position. For the shock's position $x_{\rm{shock}}$, one can choose either the density peak or an inflection point ahead of the density peak, where the second derivative has its maximum. Here we chose the latter, although both methods yield essentially the same result. We obtained the location of the inflection point by an automated analysis, using as its input the density profile obtained for a wide bin width of $\Delta x = 3b$. In this analysis, a Savitzky-Golay fit yielded the second derivative profile, for each frame. The shock position $x_{\rm{shock}}$ obtained this way was plotted as a function of time, for approximately 100 video frames, resulting in data points that fit a straight line very well. The slope of that straight line yielded the shock speed $v_{\rm{shock}}$.}. We found that the shock speed was nearly the same, for a solid as compared to a liquid, as listed in Table~\ref{tWidthParam}.

\begin{table*}[]
	\caption{Experimental conditions and measurements of shocks. Runs were performed in pairs, where the letters ``S'' and ``L'' denote solid and liquid conditions, where the liquid conditions were attained by rastered laser heating. Aside from the use of laser heating, the conditions were the same for each pair of runs.  The runs denoted 1S, 2S and 3S are the same as Runs 1, 2 and 3 of~Ref.~\cite{kananovich2020shocks}. The shock width was obtained as in Fig.~\ref{fWidth}, with a measurement uncertainty $\pm 0.1$~mm, which is the bin size. The shock speed was measured separately for the solid and liquid runs, but the exciter speed was identical for the solid and liquid runs. The measurement uncertainty for the shock speed was~$\ll 1$~mm/s.}
		\begin{ruledtabular}
	\begin{tabular}{p{2.6in} c p{0.7in} p{0.7in} p{0.7in} }

		& Symbol                   & Runs 1S, 1L & Runs 2S, 2L & Runs 3S, 3L \\ \hline
		\multicolumn{5}{ l }{\textbf{Conditions}}                                                                                                                  \\ \hline
		exciter speed (mm/s)                                                         & $v_{\rm{exciter}}$                 & 44.5        & 50.8        & 57.2        \\ 
		exciter Mach number                                                          & $M_{\rm{exciter}} = v_{\rm{exciter}} / c_l$ & 2.8         & 3.2         & 3.6         \\ \hline 
		\multicolumn{5}{ l }{\textbf{Measurements}}                                                                                                                \\ \hline
		shock   speed in \emph{solid} (mm/s)                                                & $v_{\rm{shock}}$                   & 37.0			          & 45.2            & 44.0		
		                              \\ 
		shock   speed in \emph{liquid} (mm/s)                                               & $v_{\rm{shock}}$                    & 38.2	       &47.0          & 44.1
                              \\ 
		shock Mach number   in \emph{solid}                                                 & $M_{\rm{shock}} = v_{\rm{shock}} / {c_l}$     & 2.3         & 2.8         & 2.7         \\ 
		shock   Mach number in \emph{liquid}                                                & $M_{\rm{shock}} = v_{\rm{shock}} / {c_l}$     & 2.4         & 2.9         & 2.8         \\ 
		shock width in \emph{solid} (mm) 													 & $\delta$                        & 2.7         & 2.2         & 1.6         \\ 
		shock width   in \emph{liquid} (mm)                                                 & $\delta$                        & 1.8         & 1.5         & 1.4         \\ 
		dimensionless shock width in \emph{solid}                                           & $\delta / b$                    & 5.90        & 4.71        & 3.83        \\ 
		dimensionless shock width in \emph{liquid}                                          & $\delta / b$                    & 3.92        & 3.24        & 3.00        \\ 
	\end{tabular}
		\end{ruledtabular}
	\label{tWidthParam}
\end{table*}

\begin{figure}[ht]
	\centering
	\includegraphics[width=1.0\columnwidth]{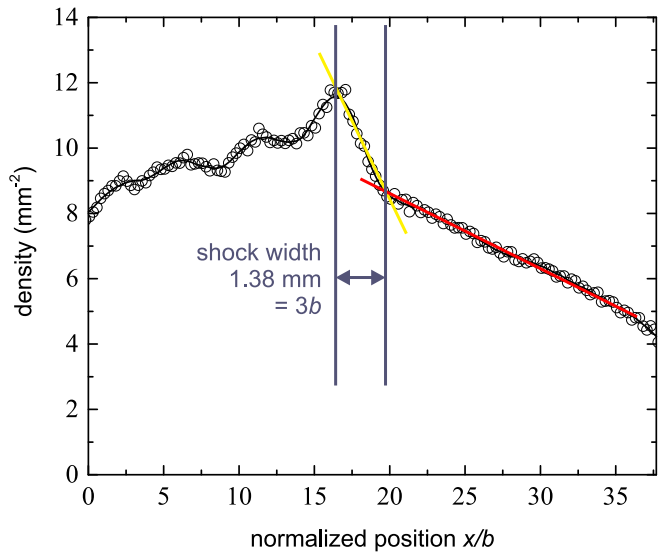}
	\caption{Density profile from the Galilean-transformation method, combining data from 25 consecutive video frames, to yield the data points shown here. A narrow bin width $\Delta x = 0.25\,b$ was used. The smooth curve was obtained using a Savitzky-Golay filter. The shock width was measured between two points on the profile: on the left we used the peak density in the smooth curve, and on the right the inflection point obtained as the intersection of two asymptotes. These profiles, along with those in Figs.~\ref{fWidthBinning} and \ref{fWidthProfileExample025b}, were from Run 3L with liquid conditions.}
	\label{fWidth} 
\end{figure}

An example density profile is shown in Fig.~\ref{fWidth}, for the Galilean-transformation method described above, combining data from 25 frames to yield the data points. Also shown is a smooth curve, obtained from the data points by using a Savitzky-Golay filter.

We measured the shock width $\delta$ as the difference in the $x$ positions of two features on the profile. One feature is the density profile's peak, which in the example of Fig.~\ref{fWidth} is at $x/b = 16.57$. The other feature is the inflection point at the front of the shock. For this purpose, the inflection point was identified as the intersection of two asymptotes, which in Fig.~\ref{fWidth} is at $x/b = 19.57$. Both of these values were obtained from the Savitzky-Golay smoothed curve for the density profile.

\section{Results}

\subsection{Features of the density profile}

From the density profiles, we found that in our 2D dusty plasma the shock width generally varied from 3 to 6 lattice constants,  \textit{i.e.}, $3 < \delta /b < 6$, as shown in Table~\ref{tWidthParam}. 

Such a shock width is generally in the range of previous measurements of shocks in 3D dusty plasmas. Usachev~\textit{et~al.}~\cite{usachev2014externally}  described the shock width as being about an interparticle distance in the undisturbed microparticle cloud~\cite{usachev2014externally}. They reported shock widths as small as 0.2~mm, which was about the same as the 0.18~mm Wigner-Seitz radius based on their reported value of dust number density. Under different conditions, Jaiswal~\textit{et al.}~\cite{jaiswal2016experimental} reported a shock width varying from 1.5 to 3.5~mm, an order of magnitude greater than their Wigner-Seitz radius of 0.13~mm. We believe that our measurements were made with a finer spatial resolution than in those previous experiments because of our Galilean-transformation method that combines data from 25 frames.

One of our chief results is a comparison of liquid and solid conditions. Our finding is that \emph{the shock width is less in a liquid than in a solid}. Although the shock speeds are nearly the same in liquids and solids, the shock width is not. These results are presented in Table~\ref{tWidthParam}, where we see that for each pair of runs, the shock width is less in a liquid.

\begin{figure}[ht]
	\centering
	\includegraphics[width=1.0\columnwidth]{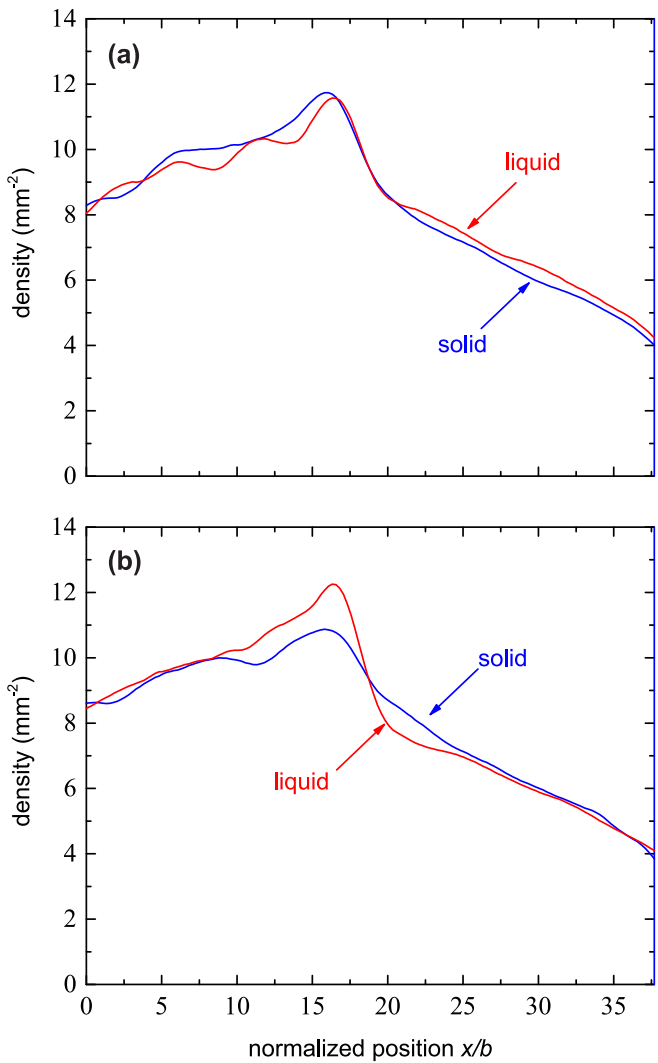}
	\caption{Comparison of density profiles for solid and liquid conditions, under shock compression. In the liquid, the gradient is higher, and accordingly the shock width is less. Data shown are from Runs 3S and 3L in (a), and 2S and 2L in (b). We also observe oscillations in the density profile, behind the shock $0 < x/b < 14$. These oscillations appear most conspicuously in the liquid in (a) and the solid in (b). Detection of these oscillations was made possible by our improved spatial resolution, using the Galilean transformation combining data from 25 frames, as in Fig.~\ref{fWidth}.}
	\label{fWidthLiquid} 
\end{figure}

This difference in shock width, for liquids vs solids, is also visible in the density profiles shown in Fig.~\ref{fWidthLiquid}.  There we see that in a liquid, the density gradient is greater, and correspondingly the shock is thinner. 

In addition to the shock width, we note another feature in the density profile: a compressional oscillation behind the shock. We observed this oscillation, which has a wavelength of about 5 to 10 lattice constants, not only in our solids, but also our liquids. We cannot definitively explain these oscillations. We can mention that oscillations have been observed behind shocks in numerical simulations of 2D Yukawa solids for both Lenard-Jones~\cite{zhakhovskii1999shock} and Yukawa~\cite{marciante2017thermodynamic} crystals. In those simulations, oscillations were reported only for solids, not liquids~\cite{anisimov1997shock,zhakhovskii1999shock}, suggesting that either the full nature of our oscillations is not captured by the simulations, or the oscillations in our experiment arise from a mechanism different from that in the simulations.

\subsection{Test of hydrodynamic description}

We test the hydrodynamic description for shocks in a strongly coupled dusty plasma by comparing our measured shock width to the value $\delta_h$ predicted hydrodynamically.  For gas dynamics, the finite shock width is often attributed to viscous dissipation. Accordingly, the shock width in this hydrodynamic description depends on the viscosity, with a predicted value~\cite{granger1985fluidshockwidth}

\begin{equation}
\label{eqWidth}
\delta_h = \frac{\nu}{V_b}.
\end{equation}
Here, $\nu$ is the kinematic shear viscosity of the substance, and $V_b$ is the bulk speed behind the shock.

This theory was originally derived for shocks in gases. For molecular liquids, this the hydrodynamic prediction of Eq.~(\ref{eqWidth}) for the shock width was found to give good agreement with simulations~\cite{hoover1979structure}, even though the sample was a liquid rather than a gas. We ask here whether a similar agreement with the hydrodynamic prediction can be attained in our sample, a strongly coupled dusty plasma.

We can estimate the value of the shock width $\delta_h$ as predicted by the hydrodynamic model, for our experimental conditions. The kinematic viscosity, as described in Sec.~\ref{sec_apparatus}, is estimated as $\nu = 1.86$~mm${}^2$/s. The bulk speed was $V_b = 18.9$~mm/s, measured in Run~3L of our experiment as the overall speed of microparticles behind the shock front. Using these two values in Eq.~(\ref{eqWidth}), we estimate $\delta_h \approx 0.098$~mm, as the prediction of the hydrodynamic description.

This predicted value of $\delta_h \approx 0.098$~mm is an order of magnitude smaller than the shock width in the range $1.4 < \delta < 1.8$~mm that we obtained experimentally, as in Table~\ref{tWidthParam}. This disagreement indicates \emph{a failure of the hydrodynamic approach, within the high-gradient region of our shock}.

Moreover, we note an underlying reason for this disagreement, between the hydrodynamic prediction and our experiment. The hydrodynamic prediction of 0.098~mm is less than the interparticle spacing. Hydrodynamics in general requires gradients to have a scale length larger than the interparticle spacing, so that the discreteness of particles within the fluid can be ignored.
  
\section{Conclusions}   

Using data from a recent experiment in a two-dimensional strongly coupled dusty plasma~\cite{kananovich2020shocks}, we analyzed the density profiles to obtain the shock width. Previous dusty plasma experiments had reported that shock widths are generally comparable to the interparticle spacing or an order of magnitude greater. Our experiment is distinguished from these earlier works by preparing both a solid and a liquid under conditions that are generally the same, to allow a comparison. A challenge in this comparison is that the shock width is small to begin with, so that detecting a small difference requires an improved spatial resolution, which we achieved using  a new analysis method. Data were combined from multiple video images, with a Galilean transformation into the shock's inertial frame.  We used these measurements to answer two questions.

First, we asked how the shock widths compare, in a liquid vs a solid. Our experimental runs were repeated under solid conditions as well as liquid conditions, which were sustained using laser heating without changing other parameters. We found that the shock width was slightly less in a liquid than in a solid. In the solid, the shock width ranged from 4 to 6 lattice constants.

Our result that shock widths tend to be less in a liquid than in a solid, for our strongly coupled plasma experiment, is a finding that might be unexpected, based on a conceptual description. That description is that the shock layer's width is determined largely by dissipation, and the dissipation mechanisms could be different in a liquid, as compared to a solid, because of factors such as plastic deformation in a solid, viscous dissipation in a liquid, and melting which can consume energy in a substance that was a solid.

However, there is little that we found in the literature, to assess whether liquids and solids have different shock widths. We are not aware of any previous experimental comparisons of this sort, for shocks in a dusty plasma. Moreover,  for other substances, the only data we have found so far, that allow a comparison, are from separate papers for  simulations of a liquid~\cite{anisimov1997shock} and a solid~\cite{zhakhovskii1999shock} that obey a Lenard-Jones potential. Further work would be required to explain the quantitative difference in the shock width,  solid vs liquid, that we observed for a dusty plasma, and to determine whether this tendency applies to other substances as well.

Second, we asked whether the  hydrodynamic description can accurately describe a shock in a strongly coupled dusty plasma. As with the first question, we relied on measurements of shock widths, which were made possible by our improved spatial resolution. We found that the hydrodynamic description fails, for the shocks in our strongly coupled dusty plasma. This conclusion is based on a discrepancy, greater than tenfold in magnitude, between our measured shock width and the value predicted by the hydrodynamic model. Although a hydrodynamic approach is useful for describing other phenomena in a strongly coupled dusty plasma~\cite{donko2006shear,feng2012energy,feng2013longitudinal,haralson2016temperature,haralson2017overestimation,wong2017strongly,haralson2018dusty,belousov2016skewness,wong2018multiple}, it is not useful within the high-gradient region of a shock front.

As an additional result, we detected compressional oscillations located behind the moving shock. These oscillations were observed not only in solids but also in liquids. They will require further study to determine their cause, and to assess whether they are related to oscillations observed in molecular-dynamics simulations of 2D Yukawa systems~\cite{marciante2017thermodynamic}.

\begin{acknowledgments}
This work was supported by U.S. Department of Energy grant DE-SG0014566, the Army Research Office under MURI Grant W911NF-18-1-0240, and NASA-JPL subcontract 1573629.
\end{acknowledgments}

\end{document}